\documentclass[prb,twocolumn,aps,superscriptaddress,showpacs,floatfix]{revtex4}
\usepackage{amsmath}
\usepackage{graphicx}
\usepackage{color}
\usepackage{tikz}
\begin{document}
\title{Majorana fermions at the edge of superconducting islands}

\author{R.S. Akzyanov}
\affiliation{Moscow Institute of Physics and Technology, Dolgoprudny,
Moscow Region, 141700 Russia}
\affiliation{Institute for Theoretical and Applied Electrodynamics, Russian
Academy of Sciences, Moscow, 125412 Russia}
\affiliation{All-Russia Research Institute of Automatics, Moscow, 127055
Russia
}

\author{A.L. Rakhmanov}
\affiliation{Moscow Institute of Physics and Technology, Dolgoprudny,
Moscow Region, 141700 Russia}
\affiliation{Institute for Theoretical and Applied Electrodynamics, Russian
Academy of Sciences, Moscow, 125412 Russia}
\affiliation{All-Russia Research Institute of Automatics, Moscow, 127055
Russia }

\affiliation{CEMS, RIKEN, Saitama, 351-0198, Japan}

\author{A.V. Rozhkov}
\affiliation{Moscow Institute of Physics and Technology, Dolgoprudny,
Moscow Region, 141700 Russia}
\affiliation{Institute for Theoretical and Applied Electrodynamics, Russian
Academy of Sciences, Moscow, 125412 Russia}
\affiliation{CEMS, RIKEN, Saitama, 351-0198, Japan}

\author{Franco Nori}
\affiliation{CEMS, RIKEN, Saitama, 351-0198, Japan}
\affiliation{Department of Physics, University of Michigan, Ann Arbor, MI
48109-1040, USA}

\begin{abstract}
We investigate the properties of electron states localized at the edge
of a superconducting island placed on the surface of a topological
insulator in a magnetic field. In such systems, Majorana fermions emerge if an odd number of vortices (or odd multivortex vorticity) is hosted
by the island; otherwise, no Majorana states exist. Majorana states
emerge in pairs: one state is localized near the vortex core, and another at
the island edge. We analyze in detail the robustness of Majorana
fermions at the edge of the island threaded by a single vortex. If the
system parameters are optimized, the energy gap between the Majorana
fermion and the first excited state at the edge is of the order of the
superconducting gap induced on the surface of the topological insulator.
The stability of the Majorana fermion state against a variation of the
gate voltage and its sensitivity to the magnetic field allows one to
experimentally distinguish the edge Majorana fermion from conventional
Dirac fermions.
\end{abstract}
\date{\today}


\maketitle

\section{Introduction}

The possible realization of Majorana fermion states in condensed
matter physics is attracting considerable interest in recent
years~\cite{Wilczek,Wilczek1,Alicea,Leijnse,Stanescu_Tewari,Franz,Beenakker,Franz_revmodphys}.
This is partly due to the non-abelian anyonic
statistics of Majorana fermions, allowing the realization of
topologically-protected quantum gates~\cite{tqc}.
Topological quantum computation requires the braiding of
anyons~\cite{kitaev_braid}.
Majorana braiding might be realized by the controllable manipulation of
the pairwise interaction between separate Majorana
fermions~\cite{Clarke_braid,Akhmerov_anyon_braid,Beenakker_braid}.
The decoherence caused by the tunneling between Majorana fermions sets an
upper limit of the time on the elementary operation, while the energy of
the excited states determines the lower time limit of the elementary
operation~\cite{splitMajorana_fermion}.
Many attempts have been performed to find Majorana fermions in different
systems. Recently, possible observations of Majorana fermions in
quantum
nanowires~\cite{exp1,exp2}
and
atom chains~\cite{atomic_chains}
were reported.

The interface between a topological insulator and an $s$-wave
superconductor (SC) is a promising system for the possible realization of
Majorana fermions.~\cite{fu_kane_device,sau_robustness,feigel1,AL,feigel2}
Such proximity-induced superconductivity is a mixture of
$s$ and $p$-wave correlations.~\cite{fu_kane_device,Golubov} Being
topologically equivalent to the $p$-wave superconductivity, it supports
Majorana fermions.~\cite{Read_Green} However, specific features of this superconducting
state, in particular, quasiparticle linear dispersion on the surface of the
topological insulator, are of importance for the structure and robustness
of the Majorana state and requires adequate analysis.
Majorana fermions may emerge at the vortex core in the proximity-induced
superconducting region on the surface of the topological insulator.
However, the minigap separating the Majorana fermion and the Caroli-de
Gennes-Matricon (CdGM) levels in the Abrikosov vortex core~\cite{zhen,Volovik_vortex} is very small (about $10^{-3}$~K). In order to increase the robustness of the Majorana state, Ref.~\onlinecite{AL} proposed to
make a hole in the superconducting layer to pin the vortex and to remove the
CdGM levels. These ideas were further elaborated in
Ref.~\onlinecite{me}.

Majorana fermions could also localize near boundaries of the superconducting region. However, such states have attracted much less attention than the Majorana fermions near the vortex core. In Refs.~\onlinecite{edge_stone, Alicea} it has been shown that the edge Majorana fermion localizes at the interface between $p$-wave superconductor and a topologically trivial insulator if odd number of vortices penetrates superconductor.
Edge Majorana modes were studied
theoretically~\cite{EPL_braid}
in a finite-size heterostructure made of a SC, a ferromagnetic insulator,
and semiconductor with strong spin-orbit coupling. It has been argued in
Refs.~\onlinecite{Tiwari1,Tiwari2}
that Majorana fermions can arise at the edge of a semi-finite SC placed on
the surface of a topological insulator in a magnetic field.

Significant progress with making superconducting islands on the surface of insulators or metals has been achieved in recent years. Observations of vortices and multivortices in the Pb superconducting islands~\cite{Rodichev,Void,Stolyarov} and regular structures of the Nb superconducting islands~\cite{Mason} have been reported. These systems are of special interest for possible implementation of the Majorana fermion surface codes for topological quantum computations~\cite{MFC,Fu_MFC}.

In our previous works~\cite{AL,me} we investigated the Majorana fermion in the core of the vortex pinned by a hollow channel in $s$-wave superconductor placed on the top of topological insulator. This channel removes CdGM levels in the core of the vortex in $s$-wave superconductor making Majorana fermion robust~\cite{AL}.

\begin{figure}[t!]
\center
\includegraphics[width=1\columnwidth]{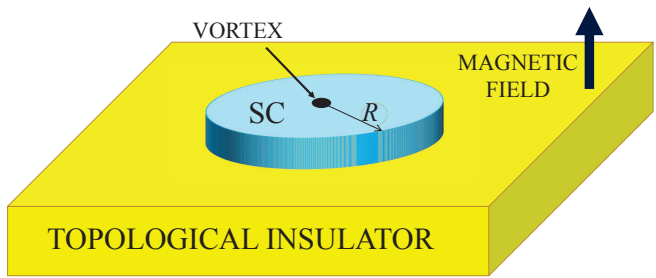}
\caption{(Color online) Proposed experimental setup for the detection of
edge Majorana fermions. A superconducting (SC) island of radius $R$ (blue) is
placed on top of a topological insulator (yellow). An
external magnetic field is perpendicular to the interface. The black circle at
the center of the superconducting island represents a vortex.}
\label{stm_exp}
\end{figure}

In this paper we consider different system: thin cylindrical $s$-wave superconducting island of radius $R$ placed on the infinite surface of a topological insulator in a transverse magnetic field $B$ (see Fig.~\ref{stm_exp}). If the island traps a vortex, two Majorana fermions are induced, one at the vortex core in the center of the island and the other at the island edge.

This paper studies the Majorana states localized near the edge of a
superconducting island. Edge Majorana states exist only if the vorticity $l$
threading the island is odd, and disappear if $l$ is even or zero.
The energy splitting between edge and vortex core Majorana fermions decays
exponentially, when increasing the radius of the island. We will
demonstrate that the edge Majorana fermion is robust: the gap between the
Majorana fermion state and the edge excited state is of the order of the
induced superconducting gap. This makes the edge Majorana state promising
for experimental observation, and, possibly, manipulation.

The presentation below is organized as follows. In
Sec.~\ref{bdg}
we derive the Bogolyubov-de Gennes equations for our system.
In Sec.~\ref{zero_modes} we investigate modes with zero energy for different number of vortices in the island. In
Sec.~\ref{sec_single_vortex}
the system with a single vortex is considered. In
Sec.~\ref{sec::discussion}
the obtained results are discussed, and conclusions are presented.

\section{Bogolyubov-de Gennes equations}\label{bdg}

\subsection{Microscopic model}

Our system is shown in
Fig.~\ref{stm_exp}.
It consists of a thin cylindrical superconducting island of radius $R$ placed on
the surface of a topological insulator. The entire heterostructure is in a
magnetic field perpendicular to the topological insulator surface.

To study the microscopic properties of such a system we will use the
formalism of
Ref.~\onlinecite{sau_robustness}.
The Hamiltonian can be written as
\begin{equation}\label{H}
 H=H_{\rm TI}+H_{\rm SC}+{T} +{T}^\dagger,
 \end{equation}
where
$H_{\rm TI}$
describes the topological insulator (TI),
$H_{\rm SC}$
describes the s-wave superconductor (SC). The term $T$ accounts for the tunneling
from the TI to the SC, while
$T^\dagger$
represents the opposite processes: tunneling from the SC to the TI.
The corresponding Bogolyubov-de Gennes equations  are
(after setting $\hbar = e = c = 1$)
\begin{equation}\label{H_ti}
H_{\rm TI}{\psi}_{\rm TI}+T^\dagger {\psi}_{\rm SC}=\omega{\psi}_{\rm TI},
\end{equation}
\begin{equation}\label{H_sc}
 H_{\rm SC}{\psi}_{\rm SC}+T {\psi}_{\rm TI}=\omega{\psi}_{\rm SC}.
 \end{equation}
The terms
$H_{\rm TI}$,
and
$H_{\rm SC}$
can be written as
$4\times4$
matrices in the Nambu basis
\begin{eqnarray}\label{Nambu}
\nonumber
H_{\rm TI} &=&
[iv(\sigma\cdot\nabla_r)-U(\textbf{r})]\tau_z+v(\sigma\cdot \textbf{A})\tau_0,
\\
H_{\rm SC} &=&
	\!-\!\left[
		E_{\rm F}\! +\!
	\frac {({\nabla}_{\textbf{R}} - i {\bf A} \tau_z)^2}
			{2m}\right]\tau_z \!
\\
\nonumber
&+&
		\Delta'(\textbf{R})\tau_x\!+\!\Delta''(\textbf{R})\tau_y,
\end{eqnarray}
and
$T=\tau_z{\cal T}(\mathbf{R-r})$.
In these equations,
$\textbf{R}=(x,y,z)$
is a point in the bulk of the SC,
$\textbf{r}=(x,y)$
is a point on the surface of the TI,
$\sigma_j,\tau_j$
are the spin and charge Pauli matrices,
$\Delta', \Delta''$
are the real and imaginary parts of the order parameter in the s-wave superconductor, $v$ is
the Fermi velocity of the electrons on the surface of the TI,
$E_{\rm F}$
is the Fermi energy in the SC,
$U(r)$
is a gate voltage applied to control the Fermi level in the TI, and
$\textbf{A}$
is the vector potential of the magnetic field. The wave functions
$\psi_{\rm TI,\,SC}$
are four-component spinors
\begin{equation}
\psi_{\rm TI,\,SC}=[u_\uparrow,u_\downarrow,v_\downarrow,-v_\uparrow]^T.
\end{equation}
It is easy to check that $H$ satisfies the following charge-conjugation
symmetry condition
\begin{eqnarray}\label{charge_conj}
\nonumber
\{ H, \Xi \}=0,\\
\Xi = \sigma_y \tau_y K,
\end{eqnarray}
where $K$ is the complex conjugation operator. Consequently, for every
eigenstate $\psi$ of the Hamiltonian $H$ with a non-zero eigenenergy
$\omega \ne 0$, an eigenstate $\Xi \psi$ with eigenenergy $-\omega$ exists.
The latter symmetry is robust: small disorder does not destroy this
property.

\subsection{Effective Hamiltonian}

In this subsection we will derive the effective description for the wave
function of the electrons on the surface of the TI. To this end we exclude
$\psi_{\rm SC}$
from
Eqs.~\eqref{H_ti}
and~\eqref{H_sc}
to obtain
\begin{eqnarray}
\label{bdg_eq}
(H_{\rm TI} + \Sigma )\psi_{\rm TI}=\omega\psi_{\rm TI},
\\
\Sigma = T^\dagger(\omega - H_{\rm SC})^{-1}T.
\end{eqnarray}
We are interested in the bound states with energies lying within the
superconducting energy gap, $|\omega|<|\Delta|$. In this case, the
self-energy matrix $\Sigma$ was calculated in
Refs.~\onlinecite{sau_robustness,AL}.
For the low-lying electron states with wave vectors
${\bf k}$
near the Dirac cone apex
${\bf M}$
in the Brillouin zone of the TI,
${\bf k}\approx{\bf M}$,
the self-energy is equal to
\begin{equation}\label{Sigma}
\Sigma_{{\bf M}, \omega}
=
\lambda\frac{\Delta \tau_x - \omega \tau_0}
            {\sqrt{|\Delta|^2 - \omega^2}}-\delta U \tau_z,
\end{equation}
where $\tau_0$ is the $2\times2$ unit matrix.
This expression for $\Sigma$ is independent of
$\bf A$.
Such an approximation is valid for weak magnetic fields, which is assumed to
be the case everywhere in this paper. The parameter $\lambda$ has the
dimension of energy. It characterizes the transparency of the interface
between the TI and the SC: when
$\lambda \sim  E_{\rm F}$
($\lambda \ll  E_{\rm F}$),
the barrier is transparent (non-transparent). The value
$\delta U=O(\lambda)$
is the shift of the chemical potential of the TI due to doping by the
carriers coming from the SC.

Using
Eq.~(\ref{Sigma})
we can rewrite the Bogolyubov-de Gennes
equation~(\ref{bdg_eq})
in the form
\begin{equation}\label{bdg2}
H_{\rm eff} \psi_{\rm TI} = \omega \psi_{\rm TI},
\end{equation}
where the effective Hamiltonian
$H_{\rm eff}$
equals to
\begin{eqnarray}\label{Heff}
\nonumber
H_{\rm eff}
&=&
[i\tilde{v}(\omega)(\sigma \cdot\nabla_{\textbf{r}})-\tilde{U}(\omega)]\tau_z
+
\tilde{v}(\omega)(\sigma \cdot \textbf{A})\tau_0
\\
&+&\tilde{\Delta}'(\omega)\tau_x+\tilde{\Delta}''(\omega)\tau_y.
\end{eqnarray}
The renormalized parameters of the effective Hamiltonian are
\begin{equation}\label{renormalized_speed}
\tilde{v}(\omega)
=
\frac {v\sqrt{|\Delta|^2-\omega^2}}{\sqrt{|\Delta|^2-\omega^2}+\lambda},
\end{equation}
\begin{equation}
\tilde{U}(\omega)=\frac {(U + \delta U)
\sqrt{|\Delta|^2-\omega^2}}{\sqrt{|\Delta|^2-\omega^2}+\lambda},
\end{equation}
\begin{equation}\label{teta}
\tilde{\Delta}(\omega)=\frac {\Delta \lambda}{\sqrt{|\Delta|^2-\omega^2}+\lambda}.
\end{equation}
It is also convenient to define the renormalized coherence length:
\begin{equation}\label{renormalized_xi}
\tilde{\xi}(\omega)=\frac {v\sqrt{|\Delta|^2-\omega^2}}{\Delta\lambda}
=
\xi\frac {\sqrt{|\Delta|^2-\omega^2}}{\lambda}.
\end{equation}

If the Hamiltonian parameters are independent of
${\bf r}$,
then the eigenstates of the effective Hamiltonian obey the inequality
$|\omega| > \Delta_{\rm TI}$,
where the quantity
$\Delta_{\rm TI}$
satisfies the following equation:
\begin{equation}\label{DTI}
\frac{\Delta_{\rm TI}}{\lambda}=\sqrt{\frac{\Delta-\Delta_{\rm TI}}{\Delta+\Delta_{\rm TI}}}.
\end{equation}
The physical meaning of
$\Delta_{\rm TI}$
is the proximity-induced superconducting gap on the surface of the TI.

\subsection{Equations for the effective wave function}

Further, we assume that the island radius $R$ is much larger than the SC coherence length $\xi_{SC}$.
\begin{eqnarray}
R\gg \xi_{SC}
\end{eqnarray}
We are looking for solutions of
Eq.~(\ref{bdg2})
which correspond to bound states. The energies $\omega$ of such
eigenstates are smaller than the proximity-induced gap
$\Delta_{\rm TI}$, in
Eq.~(\ref{DTI}).
If an Abrikosov vortex with vorticity $l$ is trapped in the island, the
order parameter
$\Delta(\textbf{r})$
can be expressed as
\begin{equation}
\Delta(\textbf{r})=|\Delta(r)|\exp\left({-il\theta}\right),
\end{equation}
where $r$ and $\theta$ are the polar coordinates, and
$|\Delta(r)|\rightarrow |\Delta|$,
when
$r\gg \xi$.
If the island radius $R$ is large,
$R\gg\tilde{\xi}$,
$|\Delta(r)|$
can be approximated as
\begin{equation}\label{abs_Delta}
|\Delta(r)|=|\Delta|\Theta(R-r),
\end{equation}
where
$\Theta(r)$
is the Heaviside step function. In the geometry shown in
Fig.~\ref{stm_exp},
the vector potential can be written as
$A_z = A_r = 0$,
$A_\theta = A(r)$.
This choice of the vector potential corresponds to the magnetic field
\begin{eqnarray}
H_z = \frac{1}{r}
\frac{d(rA)}{dr}.
\end{eqnarray}

Let us introduce a spinor $F$
\begin{eqnarray}\label{F}
\nonumber
\psi_{\rm TI} &=& \exp[-i\theta(l\tau_z-\sigma_z)/2+i\mu \theta]F^{\mu}(r), \\
F^{\mu} &=& (f^{\mu}_1,f^{\mu}_2,f^{\mu}_3,-f^{\mu}_4)^T.
\end{eqnarray}
Here $\mu$ is the total angular momentum of an
eigenstate. The transformation
Eq.~(\ref{F})
is well-defined only when
\begin{equation}
\label{mu_cond}
j = \mu + \frac{l + 1}{2}
\end{equation}
is an integer. In other words, when the number of vortices $l$ is odd (even), the angular momentum $\mu$ is an integer (half-integer).

Substituting Eqs.~\eqref{Heff}, \eqref{teta}, and \eqref{F} in Eq.~(\ref{bdg2}) we derive
\begin{eqnarray}\label{final}
\nonumber
i\tilde{v}\!\left (\frac d{dr}\!+\!\frac {2\mu+l+1}{2r}\!-\!A(r)\right
)\!f^{\mu}_2\!+\!|\tilde{\Delta}|f^{\mu}_3\!-\!(\omega\!+\!\tilde{U})f^{\mu}_1\!\!=\!0,\quad
\\ \nonumber
i\tilde{v}\!\left (\frac d{dr}\!-\!\frac {2\mu+l-1}{2r}\!+\!A(r)\right
)\!f^{\mu}_1\!-\!|\tilde{\Delta}|f^{\mu}_4\!-\!(\omega\!+\!\tilde{U})f^{\mu}_2\!\!=\!0,\quad
\\
i\tilde{v}\!\left (\frac d{dr}\!+\!\frac {2\mu-l+1}{2r}\!+\!A(r)\right
)\!f^{\mu}_4\!+\!|\tilde{\Delta}|f^{\mu}_1\!-\!(\omega\!-\!\tilde{U})f^{\mu}_3\!\!=\!0,\quad
\\
\nonumber
i\tilde{v}\!\left (\frac d{dr}\!-\!\frac {2\mu-l-1}{2r}\!-\!A(r)\right
)\!f^{\mu}_3\!-\!|\tilde{\Delta}|f^{\mu}_2\!-\!(\omega\!-\!\tilde{U})f^{\mu}_4\!\!=\!0.\quad
\end{eqnarray}

Equations~(\ref{final})
have the following symmetries:
(\textit{i})
$\mu \leftrightarrow - \mu$, $f_4\leftrightarrow if_1$, $f_3
\leftrightarrow if_2$, $\tilde{U} \leftrightarrow -\tilde{U}$,
and
(\textit{ii})
$A(r) \leftrightarrow - A(r)$, $f_1\leftrightarrow f_2$, $f_3
\leftrightarrow -f_4$, $l \leftrightarrow -l$, $\mu \leftrightarrow - \mu$.
Therefore, we can consider further only $\mu,A(r)\geq0$.

\section{Zero-energy solutions}\label{zero_modes}

If
$\omega,\tilde{U}=0$,
the system of
Eqs.~\eqref{final}
decouples into two sets of equations
\begin{eqnarray}\label{1_4}
\nonumber
i\tilde{v}\!
\left(
	\frac d{dr}-\frac {2\mu+l-1}{2r}+A(r)
\right)\!f^{\mu}_1-|\tilde{\Delta}|f^{\mu}_4=0,
\\
i\tilde{v}\!
\left(
	\frac d{dr}+\frac {2\mu-l+1}{2r}+A(r)
\right)\!f^{\mu}_4+|\tilde{\Delta}|f^{\mu}_1=0,
\end{eqnarray}
and
\begin{eqnarray}\label{2_3}
\nonumber
i\tilde{v}\!\left (\frac d{dr}+\frac {2\mu+l+1}{2r} -A(r)\right )\!f^{\mu}_2+|\tilde{\Delta}|f^{\mu}_3=0, \\
i\tilde{v}\!\left (\frac d{dr}-\frac {2\mu-l-1}{2r} -A(r)\right )\!f^{\mu}_3-|\tilde{\Delta}|f^{\mu}_2=0.
\end{eqnarray}
The parameter
$|\tilde\Delta|$
is zero outside the SC island area according to
Eqs.~\eqref{abs_Delta}
and
\eqref{teta}.
It is also zero at the center of the vortex core. In this paper we are
mainly interested in the states localized at the edge of the island and the
details related to the states in the vortex core are not of importance
here. Then, for simplicity, we approximate the vortex core of the vortex with
vorticity $l$ by a cylindrical hole with a radius about the coherence
length
$\tilde{\xi}$.

\subsection{System without vortices}

Let us assume first that there are no vortices in the island. The magnetic
field localizes a zero mode near the edge of the island. However, we show
that this mode is not a robust Majorana fermion.

In the absence of vortices, the SC order parameter $\Delta$ is non-zero in the SC island and the solutions of Eqs.~\eqref{1_4} and \eqref{2_3} which are regular at $r=0$ can be expressed in terms of the modified Bessel functions $I_m(x)$
\begin{eqnarray}\label{qwer}
\nonumber
f_1\! =\! C_1\exp\left({-\int\limits_0^r A(r')dr'}\right)\!\!I_{\mu-1/2}\!\!
\left(\frac{\lambda r}{v}\right), \nonumber
\\
f_4 \!=\! -iC_1\exp\left({-\int\limits_0^r A(r')dr'}\right)\!\!I_{\mu+1/2}\!\!
\left(\frac{\lambda r}{v}\right),\nonumber
\\
\label{r>R}
f_2\! =\! C_2\exp\left({\int\limits_0^r A(r')dr'}\right)\!\!I_{\mu+1/2}\!\!
\left(\frac{\lambda r}{v}\right), \nonumber
\\
 f_3 \!=\! iC_2\exp\left({\int\limits_0^r A(r')dr'}\right)\!\!I_{\mu-1/2}\!\!
\left(\frac{\lambda r}{v}\right).
\end{eqnarray}
Outside the island, where $\tilde{\Delta}=0$, the solution of Eqs. \eqref{1_4} and \eqref{2_3} becomes
\begin{eqnarray}\label{r<R}
\nonumber
f_1=A_1\exp\left({-\int\limits_0^r A(r')dr'}\right)r^{\mu-\frac{1}2},
\\ \nonumber
f_4=A_4\exp\left({-\int\limits_0^r A(r')dr'}\right)r^{-\frac {1}2 - \mu},
\\ \nonumber
f_2=A_2\exp\left({\int\limits_0^r A(r')dr'}\right)r^{-\mu-\frac{1}2},
\\
f_3=A_3\exp\left({\int\limits_0^r A(r')dr'}\right)r^{\mu-\frac {1}2}.\quad
\end{eqnarray}
The functions
$f_2$
and
$f_3$
diverge when
$r\rightarrow +\infty$;
then,
$A_2=A_3=0$
or
$f_2=f_3=0$.
Matching solutions at
$r=R$,
one completes the derivation of the wave function.

These eigenfunctions correspond to the
$\omega = 0$
Landau level, whose states are weakly corrected to account for the presence
of the superconducting island. They are not Majorana fermion states, and a
weak perturbation of the Hamiltonian may shift their eigenenergies away
from zero value.

\subsection{System with vortices}
\label{sub::no_vort}

In this subsection we study a system with vortices. Since we are mainly
interested in edge states and a relatively small SC island, we assume
that the magnetic flux captured in the SC island forms a multivortex with
vorticity $l$ and further assume that the SC order parameter in this vortex behaves
like a Heaviside step function
$\tilde{\Delta}(r)=\tilde{\Delta}\Theta(r-\tilde{\xi})$.
This simplification neglects the CdGM states inside the superconductor, and thus,
it can significantly affect the Majorana state localized near the vortex
core. However, the edge-localized Majorana fermion is fairly insensitive to
the details at the center of the island, because its wave function decays
quickly away from the edge.

Under these assumptions, the solutions of
Eqs.~\eqref{1_4}
and~\eqref{2_3}
outside of the island $r>R$, and inside of the vortex core
$r<\tilde{\xi}$,
can be expressed as
\begin{eqnarray}
\label{r<<<R}
\nonumber
f_1=C_1\exp\left({-\int\limits_0^r A(r')dr'}\right)r^{\mu+\frac{l-1}2},
\\
f_4=C_4\exp\left({-\int\limits_0^r A(r')dr'}\right)r^{\frac {l-1}2-\mu},
\end{eqnarray}
and
\begin{eqnarray}
\nonumber
f_2=C_2\exp\left({\int\limits_0^r A(r')dr'}\right)r^{-\mu-\frac{l+1}2},
\\
f_3=C_3\exp\left({\int\limits_0^r A(r')dr'}\right)r^{\mu-\frac {l+1}2},
\nonumber
\end{eqnarray}
where $C_i$ are constants (for
$r< \tilde \xi$
these constants may be different from the constants at
$r>R$).
The functions $f_2$ and $f_3$ diverge when $r \rightarrow\infty$, and
also $f_2$ diverges when $r \rightarrow 0$. Then, as it follows from
Eqs.~\eqref{2_3}, a regular solution exists only if $f_2=f_3=0$ in the
whole space.

Inside the SC island a wave function can be expressed as a sum of two
distinctly dissimilar solutions of
Eq.~\eqref{1_4}.
A solution of the first type is localized near the vortex core:
\begin{eqnarray}
\label{eq::type1}
\nonumber
f_1\! =\! C'_2\exp\left({-\int\limits_0^r A(r')dr'}\right)r^{\frac{l}{2}}K_{\mu-1/2}\!\!
\left(\frac{\lambda r}{v}\right), \nonumber
\\
f_4 \!=\! iC'_2\exp\left({-\int\limits_0^r A(r')dr'}\right)r^{\frac{l}{2}}K_{\mu+1/2}\!\!
\left(\frac{\lambda r}{v}\right).
\end{eqnarray}
Here
$C'_{1,2}$
are constants, and
$K_n(x)$
is the modified Bessel function. Since
$K_n$
diverges at
$x=0$, the
function
$f_4$
can be normalized only if
$\mu < (l+1)/2$.
Using the symmetry between positive and negative $\mu$, one can generalize
this inequality for arbitrary $\mu$:
\begin{equation}\label{condit}
|\mu| < \frac{l+1}{2}.
\end{equation}
If $\mu$ violates this condition, then
Eq.~(\ref{eq::type1})
does not define a valid eigenstate.

Unlike
Eq.~(\ref{eq::type1}),
which describes eigenfunctions localized at
$r=0$,
a solution of the second type grows toward the edge of the island: for
$r<R$
one can write
\begin{eqnarray}
\label{eq::type2}
\nonumber
f_1\! =\! C'_1\exp\left({-\int\limits_0^r A(r')dr'}\right)r^{\frac{l}{2}}I_{\mu-1/2}\!\!
\left(\frac{\lambda r}{v}\right), \nonumber
\\
f_4 \!=\! -iC'_1\exp\left({-\int\limits_0^r A(r')dr'}\right)r^{\frac{l}{2}}I_{\mu+1/2}\!\!
\left(\frac{\lambda r}{v}\right),
\end{eqnarray}
where
$I_n$
is the modified Bessel function of the second kind. Outside the island
($r>R$)
the eigenfunction is defined by
Eq.~(\ref{r<<<R}).
Most of the wave function weight is localized away from the island center.
The value of $r$, where the weight is concentrated, grows as
$|\mu|$
increases. In the limit
$|\mu| \rightarrow \infty$,
the wave function is virtually unaffected by the presence of a
superconducting island at the origin. Thus, for large
$|\mu|$
the eigenstates described by
Eq.~(\ref{eq::type2})
become indistinguishable from the states belonging to the
$\omega = 0$
Landau level of the Dirac-Weyl electrons.

Majorana fermion states correspond to
$\mu = 0$
solutions. In finite systems these states appear in pairs. In our case, one
Majorana fermion is localized near the origin and another is localized at the
edge of the island. Inside the island their wave functions are given by
Eq.~(\ref{eq::type1})
and
Eq.~(\ref{eq::type2}).
Outside the island,
Eq.~(\ref{r<<<R})
must be used.

To demonstrate the ``Majorana nature" of the
$\mu=0$
solutions, let us calculate the first-order corrections to the energies of the
states
Eq.~(\ref{eq::type2})
caused by a non-zero, but small,
$\tilde{U}$:
\begin{eqnarray}
\label{U_correction2}
\delta \omega_{l\mu}=\tilde{U}C_{l\mu}2\pi G,\qquad \qquad \qquad\\
G\!=\!\int\limits_{r>R}\!drr^{l}Re^{-2\int\limits_0^r A(r')dr'}\left(\!\left[\frac {r}{R}\right]^{2\mu}\!-\!\left[\frac {r}{R}\!\right]^{-2\mu}\!\right) +\nonumber
\\ \nonumber\!\!
\int\limits_{r<\xi}drr^{l}Re^{-2\int\limits_0^r
A(r')dr'}
\left(
	\left[\frac{r}{\xi}\right]^{2\mu}-
	\left[\frac {r}{\xi}\right]^{-2\mu}
\right) +
\\ \nonumber
\int\limits_{\xi<r<R}\!\!\!\!\!\!\!dr
r^{l+1}
e^{-2\int\limits_0^r A(r')dr'}\!\!\!\left(\!\frac {I^2_{\mu-1/2}\!\!
\left(\frac{\lambda r}{v}\right)}{I^2_{\mu-1/2}\!\!
\left(\frac{\lambda R}{v}\right)}\!-\!\frac {I^2_{\mu+1/2}\!\!
\left(\frac{\lambda r}{v}\right)}{I^2_{\mu+1/2}\!\!
\left(\frac{\lambda R}{v}\right)}  \! \right),
\end{eqnarray}
where
$C_{l\mu}$
are wave function normalization constants. For states localized near
the vortex core of the vortex, we have a similar result with the modified Bessel
functions
$I_{\mu\pm1/2}$
replaced by the modified Bessel functions
$K_{\mu\pm1/2}$.

It follows from
Eq.~\eqref{U_correction2}
that, if
$\mu=0$,
the correction vanishes identically. Moreover, one can demonstrate that the
$\mu=0$
states are invariant under the action of the charge-conjugation operator
$\Xi$,
Eq.~\eqref{charge_conj}.
Thus, these eigenstates are topologically-protected mixtures of electron
and hole states.

For all other values of $\mu$ this correction is nonzero. Thus, the
zero-energy states with
$\mu\neq0$
are not topologically-protected from the local perturbations of the
chemical potential. In particular, the robustness of the zero-energy modes
against variations of the gate voltage may be used to distinguish Majorana
states from conventional Dirac fermions.

Finally, we would like to remind that, since $\mu$ is an integer only for odd
$l$, see
Eq.~(\ref{mu_cond}),
we must have an odd number of vortices on the island to generate Majorana
fermion states.

\section{Single vortex}\label{sec_single_vortex}

In this section we study a system with a single vortex, $l=1$, which is
the simplest in terms of a possible experimental realization. We start with the
case
$\tilde{U}=0$.
According to
Eqs.~\eqref{eq::type2}
the wave function of the zero-energy state with $\mu=0$ localized near the vortex core
is
\begin{eqnarray}\label{vortex_m_1}
\psi_v=B_ve^{-i\frac{\pi}{4}}e^{-\int \limits_0^r A(r') dr'}e^{-\int \limits_0^r \frac{\tilde{\Delta}(r')}{\tilde{v}} dr'}
\begin{bmatrix}
i\\
0\\
0\\
-1
\end{bmatrix},
\\
\nonumber
\end{eqnarray}
where we used the explicit expression for the modified Bessel function of
half-integer order.
For the edge zero-energy eigenstate, the wave function is [see
Eqs.~\eqref{eq::type1}]
\begin{eqnarray}\label{edge_m_1}
\psi_e=B_e e^{-i\frac{\pi}{4}} e^{-\int \limits_0^r A(r') dr'} e^{\int
\limits_0^r \frac{\tilde{\Delta}(r')}{\tilde{v}} dr'}
\begin{bmatrix}
1\\
0\\
0\\
-i
\end{bmatrix}.
\end{eqnarray}
Here $B_v$ and $B_e$ are normalizing coefficients, which we choose real.
Applying the particle-hole-conjugation operator $\Xi$,
we obtain that $\Xi \psi_{v(e)}=\psi_{v(e)}$ by direct calculations.
Hence, $\psi_v$ and $\psi_e$ are Majorana fermion wave functions.

\subsection{Zero-mode splitting}

The hybridization between the vortex core and the edge Majorana fermions can, in
general, split the degenerate zero level. The splitting is zero in the
case
$\tilde{U}=0$
(see also
Ref.~\onlinecite{splitMajorana_fermion}).
If
$\tilde{U}\neq0$,
the degeneracy is lifted, and Majorana fermions at the vortex core and at the
edge form the two usual Dirac states. The wave functions of these states can be
written as
$\psi_{\pm}=(\psi_v\mp i\psi_e)/2$.
These functions satisfy the particle-hole symmetry of the Bogolyubov-de
Gennes equations,
$\Xi\psi_+ = \psi_-$.
If we denote the splitting energy as
$2E_+$,
then, the wave function
$\psi_{+}$
corresponds to
$E_+$,
while
$\psi_{-}$
corresponds to
$-E_+$.
Let us assume that
$\tilde{U}$
is small. The first-order contribution to the energy splitting becomes
\begin{equation}\label{SPL}
E_+\!=\!\!\tilde{U}\langle \psi_+|\tau_z|\psi_+ \rangle.
\end{equation}
It is reasonable to assume that the applied magnetic field is smaller than
the upper critical field, that is,
$l_b\gg\tilde{\xi}$,
where
$l_b$
is the magnetic length, which satisfies
$l_b=B^{-1/2}$,
in the units used here. In this case we can neglect the effect of the
magnetic field on the wave functions near the vortex
core.~\cite{zhen}
Further, if the SC island is not large:
\begin{eqnarray}
\label{eq::size}
R\ll l_b,
\end{eqnarray}
then, using
Eqs.~\eqref{vortex_m_1},
\eqref{edge_m_1}
and
\eqref{SPL}
we derive an estimate for the energy shift
$E_+$
in the form
\begin{eqnarray}\label{splittingu0}
E_+\propto-\frac{\tilde{U}l_b}{\tilde{\xi}(0)}\exp\left[-R/\tilde{\xi}(0)\right].
\end{eqnarray}
One can now see that Majorana fermion states are robust against chemical
potential variations if the radius of the SC island is large, in the sense
that
\begin{eqnarray}
\label{eq::R>tilde_xi}
R\gg\tilde{\xi}(0) = \frac{v}{\lambda}.
\end{eqnarray}
This condition suggests that the growth of the tunneling parameter
$\lambda$ improves the isolation of the two Majorana fermions from each other,
which is a desirable property for reliable Majorana state detection.

\subsection{Excited states}

Now we calculate the energies of the excited states localized near the SC
island edge, assuming that both
Eq.~(\ref{eq::size})
and
Eq.~(\ref{eq::R>tilde_xi})
are valid, and
$\tilde{U}=0$.
It is implied for simplicity that the magnetic field penetrates the
island uniformly, so,
$A(r)=r/2l_b^2$.
In this paper, the states localized near the vortex
core~\cite{Volovik_vortex,AL,me}
are not discussed, because, when the inequality
Eq.~(\ref{eq::R>tilde_xi})
is valid, the vortex core states do not mix with the edge states, and may be
neglected.

When
$\omega \ne 0$,
$\tilde{U}=0$,
$l=1$,
and
$\tilde{\Delta} = 0$,
the system
Eq.~\eqref{final}
decouples. As a result, for
$r>R$
we have two independent systems of equations: one for the electron
components
$f_{1,2}$,
another for the hole component
$f_{3,4}$:
\begin{eqnarray}\label{ffff1234}
\nonumber
i\tilde{v}\!\left (\frac d{dr}\!+\!\frac {\mu+1}{r}\!-\!\frac {r}{2l_b^2}\right
)\!f^{\mu}_2\!\!-\!\omega\!f^{\mu}_1\!\!&=&\!0,
\\ \nonumber
i\tilde{v}\!\left (\frac d{dr}\!-\!\frac {\mu}{r}\!+\!\frac {r}{2l_b^2}\right
)\!f^{\mu}_1\!-\!\omega\!f^{\mu}_2\!\!&=&\!0,\\
i\tilde{v}\!\left (\frac d{dr}\!+\!\frac {\mu}{r}\!+\!\frac {r}{2l_b^2}\right
)\!f^{\mu}_4\!-\!\omega\!f^{\mu}_3\!\!&=&\!0, \\
\nonumber
i\tilde{v}\!\left (\frac d{dr}\!-\!\frac {\mu-1}{r}\!-\!\frac {r}{2l_b^2}\right
)\!f^{\mu}_3\!-\!\omega\!f^{\mu}_4\!\!&=&\!0.
\end{eqnarray}
Note that after the transformation
$f^{\mu}_4\rightarrow if^{-\mu}_1$
and
$f^{\mu}_3 \rightarrow if^{-\mu}_2$,
the first two equations and the second two equations exchange places. Thus,
we can solve only the equations for
$f_{1,2}$.
Substituting
$f_2$,
one derives for
$f_1$:
\begin{eqnarray}
\label{eq::f1_diff_eq}
\frac {d^2f_1^{\mu}}{dr^2}\!+\!\frac 1r \frac {d\,f_1^{\mu}}{dr}\! +\!
f_1^{\mu}\left(\frac{\mu+1}{l_b^2}\!-\!\frac {\mu^2}{r^2}-\frac{r^2}{4l_b^4}+\frac{\omega^2}{\tilde{v}^2}  \right)\!=\!0.
\end{eqnarray}
Solutions of the latter equation can be expressed in terms of the Tricomi
confluent hypergeometric functions, traditionally
denoted~\cite{abr}
as
$U(a, b, z)$
(do not confuse it with the shift of the chemical potential
$U=U(r)$,
which is a function of a single variable). As a result we have
\begin{eqnarray}\label{eq::hypergeom}
f_1^{\mu}=iAr^{\mu} \exp \left(-\frac {r^2}{4l_b^2} \right) U\!\left(-\frac{\omega^2 l_b^2}{2\tilde{v}^2},\mu+1, \frac{r^2}{2l_b^2}\right),\nonumber \\
f_2^{\mu}=\frac 12 Ar^{\mu} \exp \left(-\frac {r^2}{4l_b^2} \right) U\!\left(1-\frac{\omega^2 l_b^2}{2\tilde{v}^2},\mu+2, \frac{r^2}{2l_b^2}\right).
\end{eqnarray}
If $r\gg l_b^2 \omega/\tilde{v}$,
these wave functions decay as follows:
\begin{eqnarray}
f_1^{\mu}=iC\,\exp\left({-\frac{r^2}{4l_b^2}}\right),
\,\,
f_2^{\mu}=C\;\frac {l_b^2 \omega}{r \tilde{v}}\,\exp\left({-\frac{r^2}{4l_b^2}}\right);
\end{eqnarray}
consequently, they are normalizable. The second linear-independent solution
to
Eq.~(\ref{eq::f1_diff_eq})
diverges when
$r\rightarrow \infty$,
thus, it is not included. Near the edge
($r \approx R$)
we can approximate the functions in
Eq.~(\ref{eq::hypergeom}) as
\begin{eqnarray}
\label{bessel_r}
f_1^{\mu}=iCJ_{\mu}\left(\frac{\omega r}{v}\right),
\quad
f_2^{\mu}=CJ_{\mu+1}\left(\frac{\omega r}{v}\right),
\end{eqnarray}
if the condition
$l_b \omega /\tilde{v} \gg 1$
is satisfied. Later, we will show that this condition is similar to the
initial assumption $l_b \gg R$. The asymptotic behavior given by
Eq.~(\ref{bessel_r})
may be guessed from
Eq.~(\ref{eq::f1_diff_eq}).
Indeed, near the island edge,
$r \gtrsim R$,
the terms
$r^2/(4 l_b^4)$
and
$(\mu + 1)/l_b^2$
are much smaller than the remaining two, and may be omitted. After this
simplification,
equation~(\ref{eq::f1_diff_eq})
transforms into the Bessel equation.

In the region $r<R$, we can neglect the effect of the vector potential
since $R\ll l_b$. We introduce the following linear combinations~\cite{AL}
\begin{eqnarray}\label{Center}
 \nonumber
 X_1^{\mu}=if^{\mu}_1+f^{\mu}_4,\quad X_2^{\mu}=if^{\mu}_1-f^{\mu}_4,  \\
 Y_1^{\mu}=if^{\mu}_2+f^{\mu}_3,\quad Y_2^{\mu}=if^{\mu}_2-f^{\mu}_3,
\end{eqnarray}
where $X_{1,2}$
obey the differential equations:
\begin{eqnarray}
\frac {d^2 X_1^{\mu}}{dr^2}+ \frac 1r
\frac {dX_1^{\mu}}{dr}-\left (\frac 1{[\zeta(\omega)]^2}
+
\frac{\tilde{\Delta}}{\tilde{v}r}
-
\frac {\mu^2}{r^2} \right)X_1^{\mu} = 0,
\nonumber
\\
\frac {d^2 X_2^{\mu}}{dr^2}
+
\frac 1r \frac {dX_2^{\mu}}{dr}-\left (\frac 1{[\zeta(\omega)]^2}
-
\frac{\tilde{\Delta}}{\tilde{v}r}
+
\frac {\mu^2}{r^2} \right)X_2^{\mu} = 0,
\\
\text{where\ \ \ }
\zeta(\omega)=\frac{\tilde{v}}{\sqrt{|\tilde{\Delta}|^2-\omega^2}},
\end{eqnarray}
and
$Y_{1,2}$
can be found according to the relations:
\begin{eqnarray}\label{Ywhit}
\nonumber
Y_1^{\mu}=\frac {i\tilde{v}}{\omega}\left(\frac{dX_1^{\mu}}{dr}-\frac{\tilde{\Delta}}{\tilde{v}}X_1^{\mu}-\frac {\mu}{r}X_2^{\mu}\right), \\
Y_2^{\mu}=\frac {i\tilde{v}}{\omega} \left(\frac {dX_2^{\mu}}{dr} +\frac {\tilde{\Delta}}{\tilde{v}}X_2^{\mu}-\frac {\mu}{r}X_1^{\mu}\right).
\end{eqnarray}
As it follows from Eqs.~\eqref{final}, which are regular at $r=0$, the solutions for $X_{1,2}$ can be expressed in terms of the Whittaker functions~\cite{abr}
\begin{eqnarray}
\label{XWhit}
X^{\mu}_{1,2}
=
\frac{C_{1,2}}{\sqrt{r}} M_{\alpha_{1,2},\mu}
\left(\frac {2r}{\zeta(\omega)}\right ).
\\
\text{where\ \ \ }
\alpha_{1,2}
=
\mp \frac {|\tilde{\Delta}|}{2\sqrt {|{\tilde{\Delta}}|^2-{\omega}^2}},
\end{eqnarray}
Matching functions
$f_{i}$
at
$r=R$
and using asymptotic
Eqs.~(\ref{bessel_r}),
we obtain a transcendental equation for the eigenenergies $\omega$ of the sub-gap
excited states:
\begin{eqnarray}
\label{Energy_eq}
\left(\frac {M'_{\alpha_1,{\mu}}}{\zeta M_{\alpha_1,{\mu}}}+\frac
{M'_{\alpha_2,{\mu}}}{\zeta M_{\alpha_2,{\mu}}}-\frac
{{\mu}+1/2}R+\frac {\omega J_{\mu+1}}{\tilde{v} J_{\mu}}\right)
\nonumber
\\
\times   \left (\frac {M'_{\alpha_1,{\mu}}}{\zeta
M_{\alpha_1,{\mu}}}+\frac {M'_{\alpha_2,{\mu}}}{\zeta
M_{\alpha_2,{\mu}}}+\frac {{\mu}-1/2}R-\frac { \omega J_{\mu-1}}{\tilde{v} J_{\mu}}\right)=\nonumber\\=\left( \frac {M'_{\alpha_1,{\mu}}}{\zeta M_{\alpha_1,{\mu}}}-\frac {M'_{\alpha_2,{\mu}}}{\zeta M_{\alpha_2,{\mu}}}-\frac {\tilde{\Delta}}{\tilde{v}}\right)^2 .
\end{eqnarray}
In this expression the Whittaker functions
$M_{\alpha,\mu}(z)$
are taken at
$z=2R/\zeta(\omega)$,
the Bessel functions
$J_\alpha(z)$
at
$z=\omega R/v$,
and the prime means differentiation over $z$:
$M'_{\alpha,\mu} (z) = d M_{\alpha,\mu} (z) / dz$.
An equation similar to
Eq.~(\ref{Energy_eq})
was derived in
Ref.~\onlinecite{AL},
and later corrected in
Ref.~\onlinecite{me}.

\begin{figure}[t!]
\center
\includegraphics[width=0.95\columnwidth]{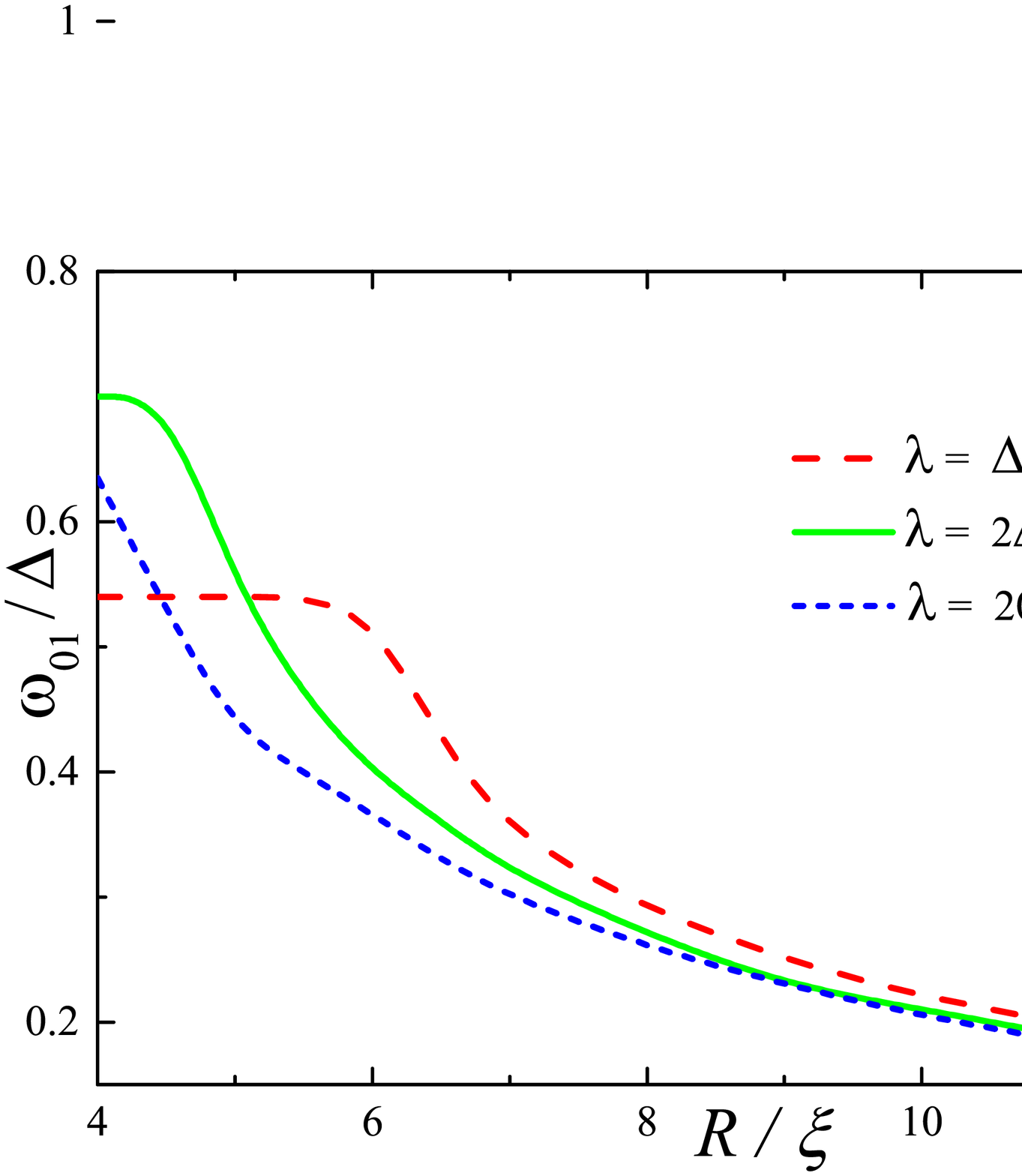}
\caption{(Color online) Normalized energy of the first excited state
($\mu=1$ and $n=0$) as a function of the normalized island radius
$R/\xi$
for different barrier transparencies $\lambda$,
$BR^2=R^2/l_b^2\ll 1$, and
according to
Eq.~\eqref{DTI},
$\lambda=20\Delta$
corresponds to
$\Delta_{\rm TI}\approx \Delta$, $\Delta_{\rm TI}\approx 0.75\Delta$,
if
$\lambda=2\Delta$;
and
$\Delta_{\rm TI}\approx 0.54\Delta$,
if
$\lambda=\Delta$.
} \label{first_ex}
\end{figure}
\begin{figure}[t!]
\center
\includegraphics[width=0.95\columnwidth]{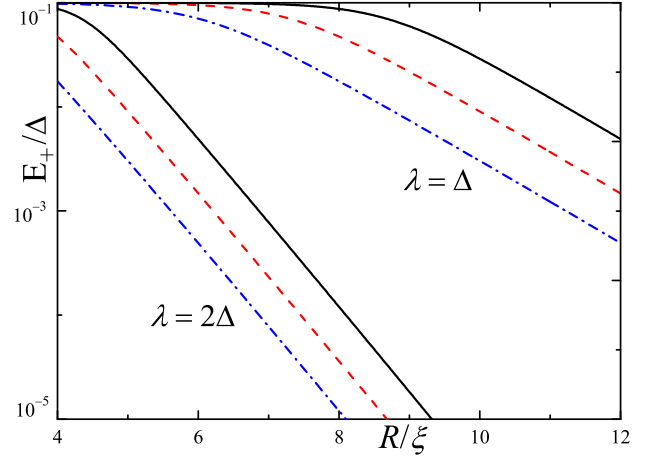}
\caption{(Color online) Normalized energy split $E_+$, Eq.~\eqref{splittingu0}, between edge and vortex Majorana fermions as a function of the normalized island radius $R/\xi$ for different barrier transparencies $\lambda=\Delta$ and $2\Delta$, shift of the Fermi level $U+\delta U=0.1\Delta$,  and different magnetic fields $\sqrt{B}R=R/l_b = 0.003$ (black) solid line, $R/l_b = 0.01$ (red) dashed line, and $R/l_b = 0.03$ (olive) dash-dotted line.}
\label{split}
\end{figure}

Each solution of
Eq.~\eqref{Energy_eq}
for the excited states can be characterized by a pair of quantum numbers:
orbital number $\mu$ and principal number $n$. A similar classification
scheme was used in
Refs.~\onlinecite{AL,zhen}.
Our numerical analysis shows that the lowest excited state, localized near
the edge, corresponds to
$\mu=1$
and
$n=0$.
The energy of the first excited state
$\omega_{01}$
as a function of $R$ is plotted in
Fig.~\ref{first_ex}
for different values of the barrier transparencies
$\lambda/\Delta$.

As one can see from this figure, the function
$\omega_{01}(R)$
decreases when the island radius $R$ increases. This function has a plateau
at low $R$ when the excited state approaches the edge of the continuous
spectrum, that is, when
$\omega_{01}$
is close to
$\Delta_{\rm TI}$.
The numerical results at large
$R/\xi$
and
$\lambda\gtrsim \Delta$
can be approximated by the
`particle-in-the-box' formula
\begin{eqnarray}\label{approx_minigap}
\omega_{01} \simeq \frac{3\pi}{4}
\Delta
\left(\frac{\xi}{R}\right)
\simeq
2.35 \frac{v}{R}.
\end{eqnarray}
It follows from the results shown in
Fig.~\ref{first_ex}
that the energy gap between the first excited state and the Majorana
fermion is of the order of the energy gap $\Delta$ in the SC if
$\lambda \gtrsim \Delta$
and
$R\leq 10\xi$.
Thus, it is not necessary to have an ideal barrier between
the TI and SC for a reasonable robustness of the Majorana fermion. For
example, if the island radius is $R=7\xi$ and $\lambda=2\Delta$, we obtain
$\omega_{10} \simeq 0.3\Delta$, which is much larger than the CdGM level
spacing $\sim \Delta^2/E_{\rm F}$ for the vortex core states~\cite{zhen} (here $E_F$
is the Fermi level of the SC). The SC island radius $R \sim 5 \xi$ is
optimal for the stability of the Majorana state if $\lambda\gtrsim \Delta$.

In Fig.~\ref{split} we show the dependence of the energy split $E_+$, Eq.~\eqref{splittingu0}, between edge and vortex Majorana fermions as a function of the island radius $R/\xi$ for different barrier transparencies and magnetic fields. Comparison of the results shown in Figs.~\ref{first_ex} and \ref{split} demonstrates that the splitting between two Majorana states is small ($E_+\ll \Delta$) and the Majorana state is rather stable ($\omega_{01}\approx 0.3-0.4\Delta$) under realistic values of parameters.

\section{Discussion}
\label{sec::discussion}

Below we discuss possible generalizations of our conclusions beyond the
constrains assumed in the previous sections, as well as connections of our
results to that of other workers.

We study the Majorana fermion near the edge of the
$s$-superconductor island on the top of the topological insulator. In our case, the edge is a
boundary between the 2D proximity-induced superconductivity and gapless
surface of the topological insulator.

One of the important parameters of the system studied is the island size.
The applied magnetic field induces a vortex in the SC island and
localizes a Majorana state near the island edge. However, if the island
radius is comparable to
the length scale
$\tilde{\xi}(0) = v/\lambda$ or $\xi_{SC}$,
the stability of the Majorana state deteriorates: first, due to the
interaction of the edge and core Majorana fermions and, second, due to
tunneling of the CdGM excited states to the edge. Therefore, in a small
island there arises a rather peculiar picture of the CdGM
states.~\cite{Melnikov}
Thus, the
condition~(\ref{eq::R>tilde_xi})
is necessary for the existence of well-defined Majorana fermions in the
system.

An island of large radius may affect the distribution of the magnetic field
in its vicinity. For our calculations we assumed that the magnetic field is
uniform. This is true if $R$ is much smaller than the effective London
penetration depth
$\lambda_{\rm eff}$
in the SC island:
$R\ll \lambda_{L\textrm{eff}}$.
However, even if this condition is violated, our results survive, provided
that the magnetic field is not too strong
\begin{equation}
\label{truecond}
BR^2=R^2/l_b^2 \ll 1,
\end{equation}
that is, the magnetic flux through the area of the island is smaller than
the flux quantum.

The latter inequality is the condition ensuring the validity of
our results. The main non-perturbative effect of the magnetic field is the
stabilization of a vortex in the island, while the inhomogeneity of the
magnetic field in the range
$r\ll l_b$
may be studied perturbatively. Indeed, the generation of Majorana
states at the vortex core and the edge depends on the parity of the vorticity
quantum $l$, and is completely unaffected by the details of the magnetic field
distribution.

Moreover, it can be shown that the relative correction to the energy of the
first excited state due to total screening of the magnetic field from the
interior of the island is of the order of
$R\tilde{\xi}/l_b^2$.
To evaluate such a correction
$\delta\omega_{01}$,
we assume that the magnetic field vanishes for
$r<R$. Then, following the procedure presented in the previous section
we find
\begin{eqnarray}
\label{correction_to_energy}
\delta \omega_{01}
=
\frac{\tilde{v}}{2l_b^2}
\frac {\int \limits_0^{R} 2 \pi r^2 dr \,\psi^{\dagger}i\sigma_y\psi}
{\int \limits_0^{\infty} 2 \pi r  dr \,\psi^{\dagger}\psi}
\,\,<\,\,
\tilde{\Delta}
\left(
	\frac {R\tilde{\xi}}{l_b^2}
\right)^2
\\
\nonumber
\sim\,\,
\omega_{01} \frac {R^2}{l_b^2}\frac {\tilde{\xi}}{R}
\,\,\ll\,\,
\omega_{01},
\end{eqnarray}
if the conditions
Eq.~\ref{truecond}
and
$\lambda\gtrsim\Delta$
are valid. For a large island
$R \gg \tilde{\xi}$,
this correction to the energy is small even if the magnetic field is quite
strong
$R \sim l_b$.

The case of stronger magnetic field or larger SC island, $R/l_b\gg 1$,
requires a separate consideration. However, as before, Majorana fermion
may exist only when the island hosts a vortex. This statement is quite
natural since the vorticity affects the quantization condition
Eq.~\eqref{mu_cond}
of $\mu$. As a result, Majorana fermions can exist only if the number
of vortices is
odd~\cite{index_theorem}.
In addition,
for each Majorana fermion, its partner must exist because the fermion parity
must be conserved. In the case of a singly-connected
SC island the only possibility is to have one Majorana fermion near the
island edge and another Majorana fermion near the vortex core, because two
Majorana fermions located in the same edge form a Dirac fermion.

In
Refs.~\onlinecite{Tiwari1,Tiwari2} the authors consider a semi-finite SC
island on top of a TI in a transverse magnetic
field. The Majorana fermion is delocalized at the edge between the SC and the TI. In such a geometry, the condition of single-valuedness of the
wave functions
Eq.~\eqref{mu_cond}
becomes a momentum quantization rule. If momentum quantization is ignored,
then the Majorana fermion can exist at the edge of the SC even in the case of
zero vorticity. This is just an unphysical artifact of using an infinite sample.

One of the major driving forces behind the development of Majorana
solid state research is the possibility of performing topological quantum
computation. To be usable in such a setup, the Majorana
fermion must be well separated from non-topological excitations.
We demonstrated that the energy of the first excited state localized at the
edge could be as large as a fraction of the superconducting gap, see
Fig.~\ref{first_ex}. The time of an elementary braiding operation must
be much less than the decoherence
time~\cite{tqc} caused by the hybridization of the edge and vortex Majorana fermions. In our system, such
decoherence time increases exponentially with the radius of the island, see Eq.~\eqref{splittingu0}.

In addition to the excitations pinned at the island edge, there are
non-Majorana states localized at
$r > R$.
These are the Landau levels which we briefly discussed in
Sec.~\ref{sub::no_vort}.
Such states with large values of $\mu$ are pushed away from the island by
the centrifugal force. As a consequence, they cannot affect the Majorana
state. At small $\mu$, however, their wave functions can reach the island
edge. Fortunately, the Landau levels are separated from each other by a gap
of the order of
$\Delta_{\rm B} \sim v/l_{\rm B}$.
Thus, at not-too-weak magnetic fields and finite $U$ these states are
shifted from the zero energy by an amount
$\sim \Delta_{\rm B}$.

Scanning tunneling microscopy (STM) might be a useful tool for
investigating Majorana
fermions~\cite{Flensberg,Ioselevich}.
In STM experiments, a Majorana fermion could manifest itself as a robust
zero-bias peak. A detailed analysis of the STM spectroscopy of the edge
Majorana fermions in the presence of vortices is done in
Ref.~\onlinecite{Tiwari2}.
The stability of the zero-bias peak should be checked against variations of
the chemical potential or the gate voltage, to distinguish the
Majorana fermions from Dirac fermions, as discussed in
Section~\ref{zero_modes}.

\begin{figure}[t!]
\center
\includegraphics[width=1\columnwidth]{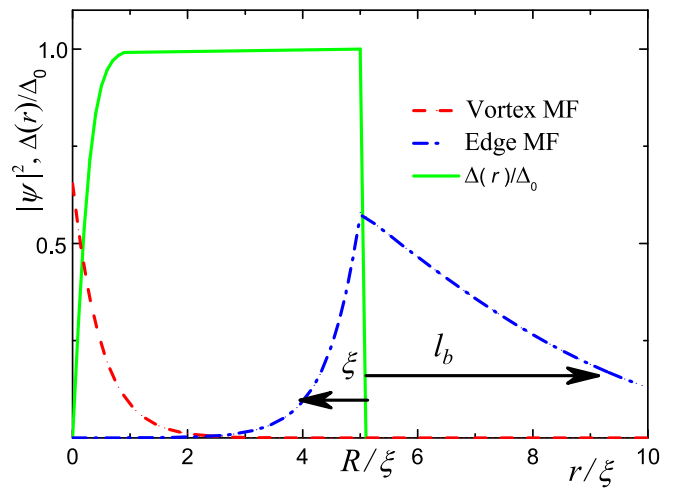}
\caption{(Color online) The value
$|\psi|^2$,
which is proportional to the local density of states, as a function of
radial position: (red) dashed line for the vortex Majorana fermion and
(blue) dashed-dotted line for the edge Majorana fermion. The local density
of states may be measured by the STM. (In the figure the value of
$|\psi|^2$
for the edge Majorana fermion is multiplied by 100.) Solid line (green)
line shows the proximity-induced superconducting order parameter
$\Delta(r)/\Delta_0$,
which vanishes when
$r>R$,
here
$\Delta_0$
is the bulk value of
$\Delta$
and
$R=l_b=5\xi$.}
\label{density}
\end{figure}

When the magnetic field is varied, the strength of the zero-bias conductance on
the island edge should oscillate when the number of vortices changes.
These oscillations could be an additional proof of the existence of
Majorana fermions. Observations of several vortices and multivortices
in superconducting Pb nano-islands were
reported.~\cite{Rodichev,Void}
Thus, measuring zero-bias peak oscillations as a function of the vorticity
is a realizable experimental task. The coordinate dependence of $|\psi|^2$ for the edge and core Majorana states are shown in Fig.~\ref{density}. The edge Majorana fermion penetrates in the island at the distance $\tilde{\xi}$ and outside the island at the distance $l_b$. With the decrease of the magnetic field (and the growth of the magnetic length) the peak value in $|\psi|^2$ for the edge Majorana state decreases and if $B \rightarrow 0$ the edge Majorana fermion becomes delocalized. Since the density of states is proportional to $|\psi|^2$, such a behavior can be observed as a zero-peak in STM measurement. Note, that the STM measurement could also reveal the zero-peak splitting due to overlapping of the wavefunctions of the edge and vortex Majorana fermions in small islands $R \sim \tilde{\xi}$ or due to the close localization of two superconducting islands.

For numerical estimates, let us consider $T_c=10$~K and $\Delta=1.76T_c \approx 2$~meV for
a BCS-type superconductor. Assuming that $\lambda=2\Delta\approx 4$~meV,
then, $\Delta=\tilde{\Delta}$ and $\xi=2\tilde{\xi}$. If we take the radius
of the island $R=7\xi$, then, the energy of the first excited state becomes
$\omega_{01}\simeq0.3\Delta\simeq 5$K. To evaluate the possible radius of the
SC island we should estimate the value of the coherence length
$\xi=v/\Delta$, which depends on the Fermi velocity on the surface of the
TI. In Ref.~\onlinecite{fermi_vel} it was reported that $v = 5.0 \times
10^7$~cm/s for the surface of Bi$_2$Se$_3$ in vacuum, then, $\xi \approx
200$~nm for the value of the gap chosen here. In Ref.~\onlinecite{Josephson_TI}
it was obtained that $v = 10^7$~cm/s on the interface between Bi$_2$Te$_3$
and a nanoribbon, and then, $\xi \approx 40$~nm. If so, then the
appropriate value
of $R$ is of the order of several hundred nm. However, it has been reported
in Ref.~\onlinecite{Landau_low} that the Fermi velocity on the surface of
Bi$_2$Te$_3$
can have a much lower value, $v = 3 \times 10^5$~cm/s, and then, $\xi
\approx 1.2$~nm and $R$ could be of the order of several nm, which might be
of the order of or lower than the coherence length $\xi_{SC}$ in the bulk
of the SC island. The case $R \lesssim \xi_{SC}$ is not optimal for the
stability of the edge Majorana fermion because CdGM states in the vortex core
in the bulk of the SC can affect the edge states.

To conclude, we studied the electronic properties of a superconducting island
in a magnetic field placed on the surface of a topological insulator.
Majorana states arise only if a vortex with odd vorticity exists
in the superconducting island. Non-topological excitations in our structure
are separated from the Majorana fermion by a significant gap, provided that
the parameters are suitably chosen. A Majorana state may be detected in
an STM experiment as a zero-bias peak, which is stable against variations of
the gate voltage. The zero-bias conductance should oscillate as a function
of the magnetic field.
We here estimate the optimal parameters for the experimental study of
Majorana fermions in our system.

\section*{Acknowledgements}

We acknowledge partial support by the Dynasty Foundation and
ICFPM (MMK), the Ministry of Education and Science of
the Russian Federation Grant No. 14Y26.31.0007, RFBR Grant No. 15-02-02128.
FN is partially supported by
the RIKEN iTHES Project, MURI Center for Dynamic Magneto-Optics
via the AFOSR award number FA9550-14-1-0040,
and a Grant-in-Aid for Scientific Research (S).

\end{document}